\documentclass[letterpaper, 10pt, conference]{ieeeconf}

\IEEEoverridecommandlockouts                              
\usepackage{epsfig} 

\title{\LARGE \bf
An Estimation of the Shortest and Largest Average Path Length in Graphs of Given Density
}

\author{L\'aszl\'o Guly\'as$^{\diamondsuit \heartsuit}$, G\'abor Horv\'ath$^{\diamondsuit \heartsuit}$, Tam\'as Cs\'eri$^{\heartsuit}$  and George Kampis$^{\diamondsuit \heartsuit}$
\thanks{$^{\diamondsuit}$ Collegium Budapest, Institute for Advanced Study, Budapest, Hungary}%
\thanks{$^{\heartsuit}$ E{\"o}tv{\"o}s University, Budapest, Hungary}%
\thanks{Correspondance to {\tt\small \{lgulyas, gkampis\}@colbud.hu}}
}

\begin{document}

\maketitle
\thispagestyle{empty}
\pagestyle{empty}

\begin{abstract}
Many real world networks (graphs) are observed to be 'small worlds', i.e., the 
average path length among nodes is small. On the other hand, it is somewhat unclear what other average path length values networks can produce. In particular, it is not known what the \textit{maximum} and the \textit{minimum} average path length values are. In this paper we provide a lower estimation for the shortest average path length ($\ell$) values in connected networks, and the largest possible average path length values in networks with given size and density. To the latter end, we construct a special family of graphs and calculate their average path lengths. We also demonstrate the correctness of our estimation by simulations.
\end{abstract}

\section{Introduction}

Many real world networks (graphs) are observed to be 'small worlds', i.e., the 
average path length ($\ell$) among nodes is small. While there exists a widely accepted formulation of
this property (i.e., $\ell$ scales with the logarithm of the number of nodes ($N$)), the intuitive meaning of 
the small world property is that the average path length is shorter "than \textit{expected}". 

On the other hand, it is somewhat unclear what average path length to expect. Observed networks tend to have short $\ell$ and many algorithmically constructed network models also share this property. In particular, Erd{\H o}s-R\'enyi (ER) random networks almost always have short path lengths. ER networks are interesting for several reasons. First, they are constructed with a minimalistic set of assumptions: given $N$ nodes, each possible link is present with a certain uniform $p$ probability. Second, their minimalistic random construction makes them an approximation of a random sample from the set of all possible graphs with the given number of nodes and the given density.

However, if random or randomly sampled graphs have short $\ell$, does this mean that all graphs share this property? No. It is easy to create counter examples, i.e., graphs with large average path lengths. Typical examples include chains or regular lattices. The question is, however, how general these counter examples are? How common they are among the set of all possible graphs? Also, chains or regular lattices assume a certain number of links for a given $N$, so they may not be constructed for a given $N$ and density ($d$, the ratio of links present among all the possible links), while the average path length naturally depends on both of parameters. This makes the interpretation of the counter examples somewhat problematic.

Unfortunately, the simple enumeration of all possible graphs with a given $N$ and $d$ is prevented by their astronomical numbers. The following formula gives this number for undirected graphs (counting all isomorphs equally).

\begin{equation}
\left|G(N,d)\right| = \left(	
\begin{array}{c}
\frac{N(N-1)}{2} \\
\\
\frac{dN(N-1)}{2}	
\end{array}
\right)
\end{equation}

This simple formula worths a moment of further consideration. While networks with $N=100$ nodes are typically considered unrealistically small and a density of $d=0.0001$ does not describe an over-populated graph, the number of possible networks with these parameters is $1.75876 \times 10^{16}$. 

Given these astronomical numbers, correct sampling would be essential to determine how average path lengths are distributed among the possible graphs. While one of the definitions of the ER networks is exactly this (\cite{ER}), it is also known that the above described mechanism to generate ER graphs yields a uniform sample if and only if the network is unconnected \cite{ER}. 

Following these considerations, in this paper we provide a lower estimation for the shortest and the largest possible average path length ($\ell$) in networks of given $N$ and $d$. To the latter, we construct a special family of graphs and calculate their average path lengths.

The paper is structured as follows. The next section provides the estimation of shortest average path length in connected networks. Section 3 introduces the family of graphs constructed to estimate the maximum average path length, while Section 4 provides analytical results for the average path length of these graphs. This is followed by a comparison of these results to the average path lengths of real-world graphs and various classic network models. The last section outlines future works and concludes the paper.

\section{Estimation of the Shortest Average Path Length in Connected Graphs}

The average path length of full graphs is obviously 1. This is because every node is connected by every other one, so every node can reach all of others on a 1-length path. If we remove only one edge from the graph we decrease the density of the graph by $\frac{2}{N(N-1)}$. The path lengths change in the following way: $\frac{N(N-1)}{2}-1$ path lengths will remain 1 and a path between a pair of nodes increases to  2. So $\ell$ is increased to:

\begin{equation}
	\frac{ \frac{N(N-1)}{2} - 1 + 2 }{ \frac{N(N-1)}{2} } = \frac{\frac{N(N-1)}{2}+1}{\frac{N(N-1)}{2}} = 1 + \frac{2}{N(N-1)}
\end{equation}

This means that $\ell$ is increased by $1-d$. Repeating the above argument until every node has only $N-1$ edge, the increase of $\ell$ in this range will be linear.

Suppose that the network has only $L$ edges. This means that there are $L$ 1-length paths. So $d\frac{N(N-1)}{2}$ paths has a length of 1 and we do not know the length of the remaining paths. But we know that their length must be at least 2, because they cannot be 0 or 1. This yields an estimation of the minimum $\ell$ of a graph with given size and density:

\begin{equation}
	\frac{\left[ d\frac{N(N-1)}{2}\right]\cdot 1 + \left[ (1-d)\frac{N(N-1)}{2}\right] \cdot 2}{\frac{N(N-1)}{2}}
\end{equation}

After some simplification this yields:

\begin{equation}
	d + (1-d) \cdot 2 = 2 - d
\end{equation}

This is the shortest possible average path length a connected graph with given size and density may have.

\section{A Family of Graphs with Large Average Path Length}

Before constructing our graphs, let us make a few observations. It is common wisdom that large $\ell$ is to be expected in 'geographical' networks, i.e., where the nodes have same phisical or topological locations and links connect closeby nodes only. Thus, chains (i.e., networks where nodes are lined up along a line and links connect nearest neighbors only) would be an ideal candidate network. The problem is, chains can only accommodate $N-1$ of links. If density dictates more than this, we need another solution. Moreover, the addition of each new link decreases the the average path length, potentially at several points. Trivially, the new link creates a path of length 1 between its two endpoints. In addition, it potentially shortens all paths going to one of the link's nodes. Furthermore, if the new link created a new shortcut, paths among two different nodes may also be shortened.

\begin{figure}[thpb]
      \centering
      \includegraphics[width=8cm]{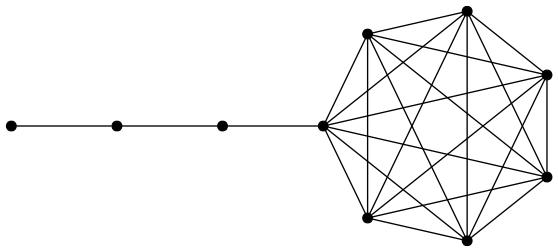}
      \caption{Example from the family of graphs constructed for maximum average path length}
      \label{example_graph}
\end{figure}

In our construction, we want to preserve as long a chain as possible. Thus, our intuition is that it is best to add additional links in a single, densely connected 'blob' (cluster) at one end of the chain. Ideally, we create a fully connected blob (a clique) and preserve the rest of the links and nodes for the chain (or tail). (See Figure \ref{example_graph}.) 

Naturally, the more links we need to place in the blob, the more nodes are needed. At the extreme, when all $N(N-1)/2$ links are present, no tail is possible. It is also clear that the length of the tail ($z$) is a non-continuous function of the link density. For example, if we have more than $(N-1)(N-2)/2+1$ edges then all nodes must belong to the blob. (For the sake of simplicity, in the following we will always assume that the blob is fully connected.)

The first step in creating a graph of this family for a given $N$ and $d$ is to determine the length of the tail ($z$). By definition, nodes in the tail may have at most 1 connection to nodes in the blob. (I.e., for clarity we define the 'last' node of the tail to be the one \textit{not} in the blob.) Thus we have $z$ nodes in the tail and $N-z$ in the blob, and since the blob is assumed to be fully connected, we get the following equation for the edges:

\begin{equation}  
	D = d \frac{N(N-1)}{2} = z + \frac{(N-z)(N-z-1)}{2},
\end{equation}

which yields the following formula for $z$

\begin{equation}\label{z}
	z = \frac{\pm \sqrt{8D - 8N + 9} + 2N - 3}{2}.
\end{equation}

A necessary condition for real roots is $8D - 8N + 9 \geq 0 \Rightarrow  D + 8/9 \geq N$. This essentially corresponds to the requirement that the graph should have enough edges to be connected, without which our construction will not work.

In order for $z$ to be meaningful, it is also necessary that $z \leq N$, which means that 
\[
  	\frac{\pm \sqrt{8D - 8N + 9} + 2N - 3}{2} \leq N
\]
\[
  	\frac{\pm \sqrt{8D - 8N + 9}}{2}  + N - 3/2 \leq N
\]
\[
  	\frac{\pm \sqrt{8D - 8N + 9}}{2} \leq 3/2
\]

Therefore, in the following we will always work with the smaller root.

\section{Calculation of the Average Path Length}

In this section we calculate the average path length in the above constructed family of graphs. First, we will calculate the sum of all shortest paths, then divide it by $N(N-1)$ (as, for the sake of simplicity, we will count all paths in both directions). 

There are three kind of nodes in our graph: i) the ones in the tail, ii) the one bridging the tail to the rest of the blob, and iii) the ones in the blob. Let's consider the $i^{th}$ node in the tail. The length of the paths connecting it to the $i-1$ nodes preceeding it in the tail is $i-j$, where $j$ is the position of the other node. Similarly, the length of the paths to the $z-i$ nodes following it in the tail is $j-i$. The path to the bridge node is of length $z-i+1$, while the $N-z-1$ nodes in the blob are reachable in $z-i+2$ steps. On the other hand, the bridge node needs $z-i+1$ steps to the $i^{th}$ node in the tail, while $1$ step to each of the $N-z-1$ nodes in the blob. For these latter, the connection to the rest of the blob, plus the bridge ($N-z-1$ nodes altogether) is of length $1$, while to the nodes in the tail it takes $z-i+2$ steps. These yield the following formula.

\[
	\mbox{SUM}(N, z) = \sum_{i=1}^{z} \sum_{j=1}^{i-1}(i-j) + 
	      \sum_{i=1}^{z} \sum_{j=i+1}^{z}(j-i) + 
	      \sum_{i=1}^{z} (z-i+1) + 
\]
\[	      
	      + \sum_{i=1}^{z} (N-z-1)(z-i+2) +
	      \sum_{i=1}^{z} (z-i+1) +
	      (N - z - 1) +
\]
\[	      
	      + (N-z-1)\left(
	      N-z-1 + \sum_{i=1}^{z} (z-i+2)
	      \right) =      
\]

\[
 				= N^2 + Nz^2 + Nz - N - \frac{2z^3}{3} - 2z^2 - \frac{4z}{3}
\]

Substituting $z$ with the smaller value in (\ref{z}) we get:

\begin{eqnarray}
			\mbox{SUM}(N, D)	& = & \frac{-{N}^{3}-6\,{N}^{2}}{3} + \nonumber \\
						            &   & \frac{\sqrt{-8\,N+8\,D+9}\,\left( 2\,N-2\,D-2\right)}{3}  + \nonumber \\
						            &   & \frac{\left( 6\,D+13\right) \,N-6\,D-6}{3}
\end{eqnarray}

or 

\begin{eqnarray}
 			\mbox{SUM}(N, d) &	= & \left( d\,{N}^{2}+\left( -d-2\right) \,N+2\right) \cdot \nonumber \\ 
 			                 &    & \frac{\sqrt{4\,d\,{N}^{2}+\left( -4\,d-8\right) \,N+9}}{3} + \nonumber \\
 			                 &    & \frac{\left( 1-3\,d\right) \,{N}^{3}+\left( 6\,d+6\right) \,{N}^{2}}{3} + \nonumber \\
 			                 &    & \frac{\left( -3\,d-13\right) \,N+6}{3}
 			\end{eqnarray}.

Therefore the path length can be given as 

\begin{eqnarray}
			l(N,D)	& = & \frac{-{N}^{3}-6\,{N}^{2}}{3\,{N}^2-3\,N} + \nonumber \\
			        &   & \frac{\sqrt{-8\,N+8\,D+9}\,\left( 2\,N-2\,D-2\right)}{3\,{N}^{2}-3\,N} + \nonumber \\
			        &   & \frac{\left( 6\,D+13\right) \,N-6\,D-6}{3\,{N}^{2}-3\,N}
\end{eqnarray}

or 

\begin{eqnarray}
      l(N,d)	& = & \left( d\,{N}^{2}+\left( -d-2\right) \,N+2\right) \cdot \nonumber \\
              &   & \frac{\sqrt{4\,d\,{N}^{2}+\left( -4\,d-8\right) \,N+9}}{3\,{N}^{2}-3\,N} + \nonumber \\
              &   & \frac{\left( 1-3\,d\right) \,{N}^{3}+\left( 6\,d+6\right) \,{N}^{2}}{3\,{N}^{2}-3\,N} + \nonumber \\
              &   & \frac{\left( -3\,d-13\right) \,N+6}{3\,{N}^{2}-3\,N}      
\end{eqnarray}.

\begin{figure*}[thpb]
      \centering

      \begin{tabular}{c c}
        \includegraphics[width=8cm]{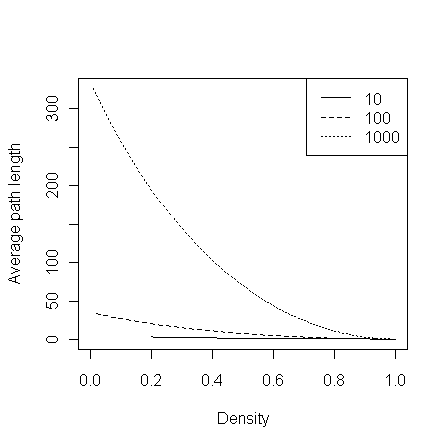} &
        \includegraphics[width=8cm]{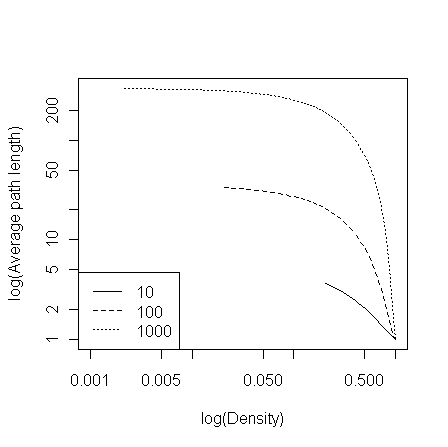} \\
      \end{tabular}

      \caption{The lower estimation of the maximum average path length value for various network sizes across the density spectrum. Note that network size determines a minimum meaningful density that drops with $N$.}
      \label{MaxAvgPlot}
   \end{figure*}

As an illustration Figure \ref{MaxAvgPlot} plots this function for various values of $N$ accross the density spectrum. 

\section{Numerical Tests of the Estimations}

\begin{figure}[thpb]
	\centering
	\includegraphics[width=8cm]{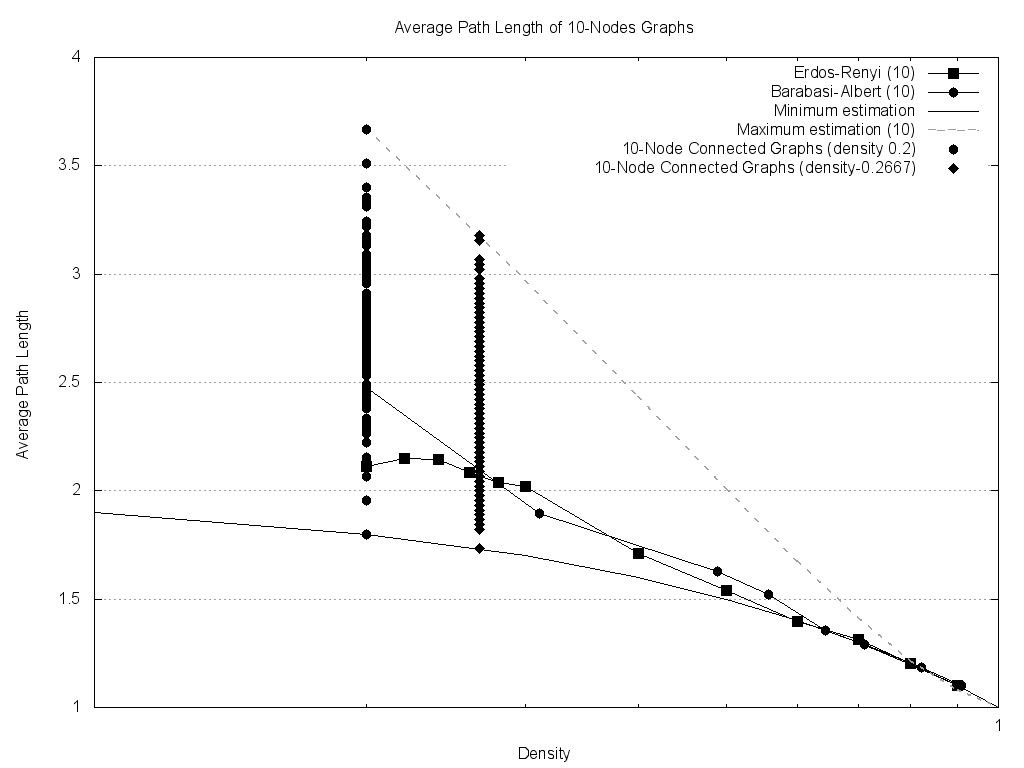}
	\caption{Average path length of 10-node graphs}
	\label{n10}
\end{figure}

\begin{figure*}[thpb]
	\centering
	\includegraphics[width=16cm]{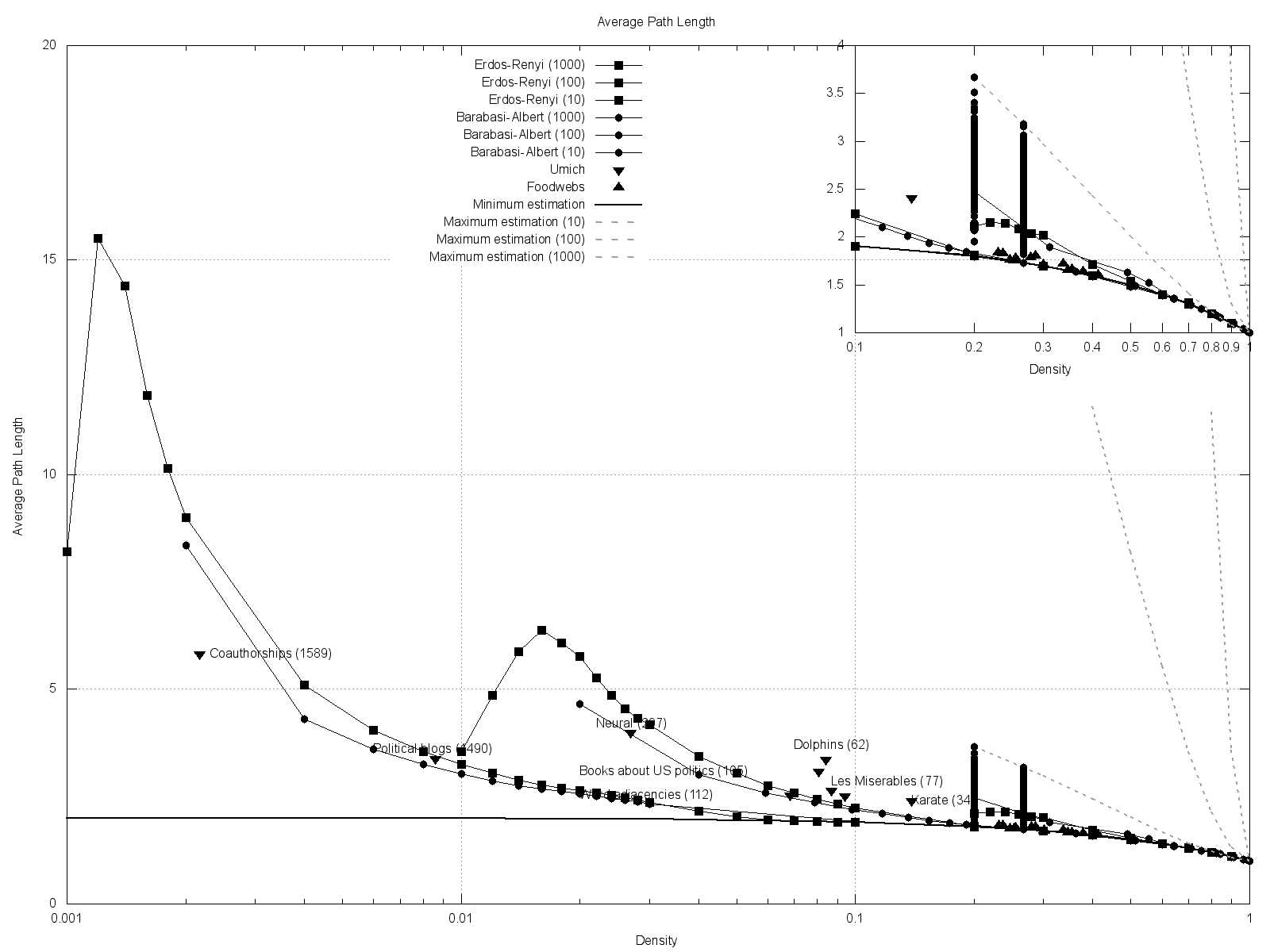}
	\caption{Average path lengths of various networks (lin-log plot)}
	\label{aplfig}
\end{figure*}

To demonstrate the correctness of our estimation, we created ER and BA networks in the entire density range for different sizes. We also enumerated all possible 10 node graphs with $d=0.2$. As shown on Figure \ref{n10} both estimations are matching with the $\ell$ of generated graphs. Figure (\ref{aplfig}) places various empirical networks on the plot as well. It is worth noting that for most densities networks are very close to the theoretical minimum. This is especially true for foodweb networks (that tend to be small that implies a relatively high density as well).

For completeness, we note that our estimation of shortest $\ell$ correct only for connected graphs. If a graph is unconnected it is possible to have shorter $\ell$.

\begin{figure*}[thbp]
	\centering
	\includegraphics[width=16cm]{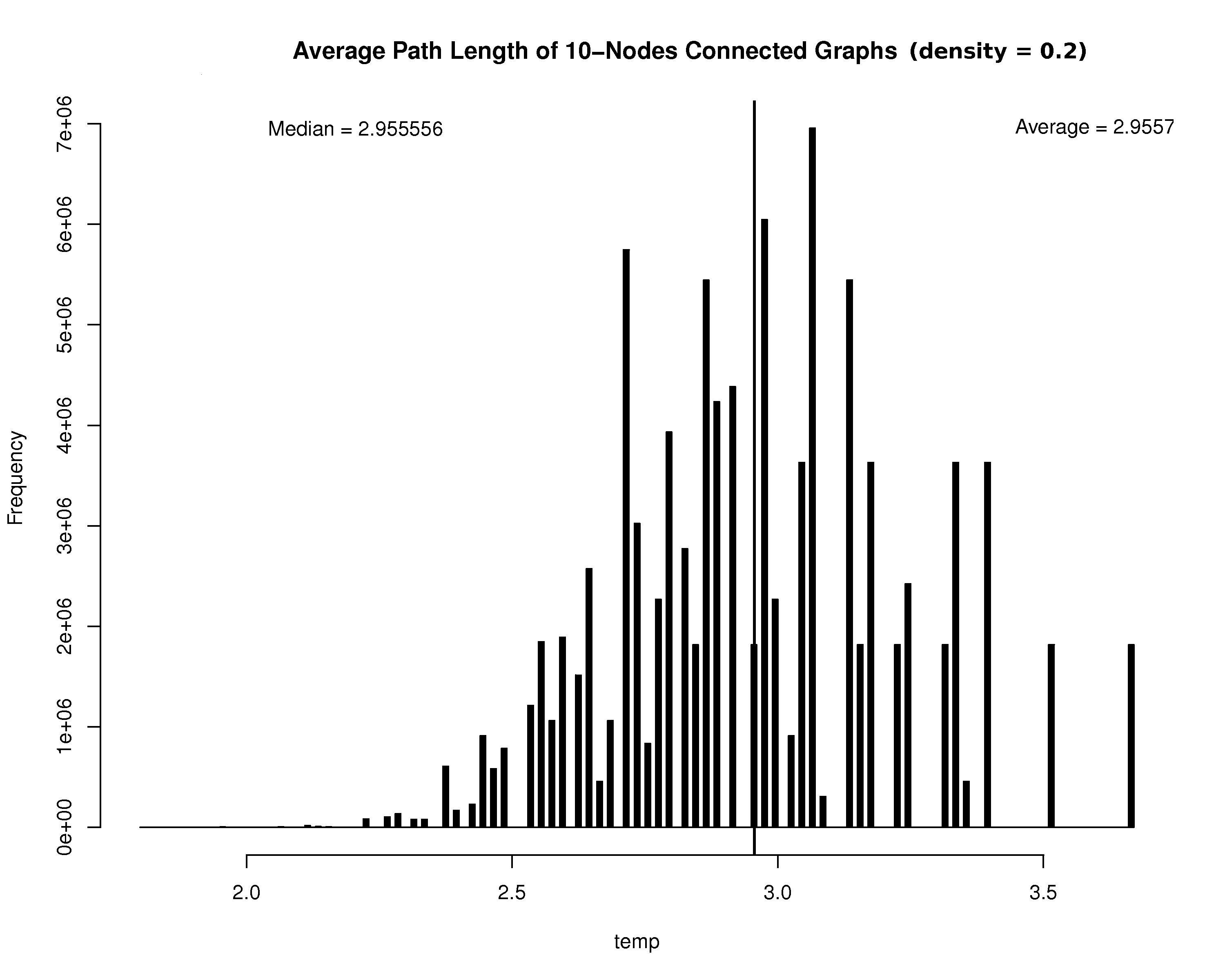}
	\caption{Distribution of average path length in 10-node graphs with d=0.2}
	\label{dist}
\end{figure*}

We generated all the 10-node graphs with density 0.2. The following figure shows the simulation result of the 10-node connected graphs. (Fig. \ref{dist}).

\section{Conclusions and Future Works}

In the paper we gave exact estimations of average path length in graphs with give size and density. 
We gave an exact estimation of shortest average path length of graphs with given size and density. Then we created a sepcial family of graphs to determine the largest $\ell$. With the help of this family we gave a lower estimation for the largest $\ell$ of graphs with given size and density. This result is important because earlier works give valid results only for certain graph types and typically only in relatively short intervals of density. We demonstrated the correctness of our estimation by generating all possible graphs with given size and density.

\section{ACKNOWLEDGEMENTS}

This research was partially supported by the Hunga-rian Government (Anyos Jedlik programme managed by the National Office for Research and Technology: TexTrend project (www.textrend.org), contract no. NKFP\_07\_A2 (2007)\/TEXTREND) and the European Union's Seventh Framework Programme: DynaNets, FET-Open project no. FET-233847 (www.dynanets.org). The supports are gratefully acknowledged.


\section{Appendix: Generating all possible graphs with given size and density}

We mentioned above in Section 1 that generating all possible graphs with given size and density gives astronomical numbers. Let \textit{N} the number of vertices in the graph and let \textit{d} the density of the graph. So the graph has $d\frac{N(N-1)}{2}$ edges (working with undirected graphs). In an undirected graph the largest number of edges can be $\frac{N(N-1)}{2}$. This means if we want to generate a graph with \textit{N} vertices and \textit{d} density, we have to select $d\frac{N(N-1)}{2}$ edges from $\frac{N(N-1)}{2}$ (see Equation 1).

If we have only 10 vertices and 12 edges (d=0.266667) this formula gives that there are 28 billion 760 million 21 thousand and 745 different graphs. Another example of how large is the number of possible graphs with given size and density is if we have 100-nodes and 495 edges (d=0.1). The formula above gives $1.3367 \cdot 10^{697}$. Put in context, in theory our universe contains "only" $10^{80}$ atoms, which is about $10^{627}$ less than the number of possible 100-nodes graphs with 0.1 density.

\begin{figure}[thbp]
	\centering
	\includegraphics[width=8cm]{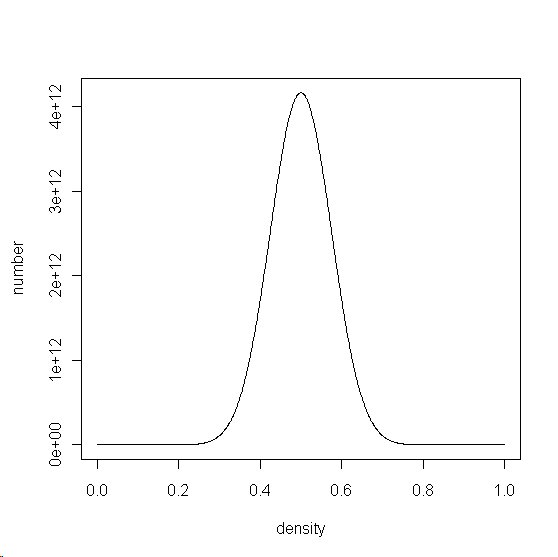}
	\caption{Number of possible graphs with given size}
	\label{num}
\end{figure}

This number highly depends on density. Although it is symmetric in the function of density because of binomial numbers. However this symmetry is surprising, because neither $\ell$ or other measured values show this symmetry. Figure \ref{num} shows the symmetry on 10-node graphs.

Sampling of $\ell$ is difficult. ER graphs are not sampling well the $\ell$ if the graph is connected. As we can see in Figure \ref{n10} the $\ell$ of ER graphs is much shorter than expected. After that, sampling $\ell$ in a non-biased way is not trivial. The sampling algorithm can be very complicated and slow. In addition, the trivial enumerating of possible graphs is also wrong, because of very large numbers of permutation. For example if we want to enumerate all of 100-nodes graphs with 0.1 density. The number of possible graphs is $1.3367 \cdot 10^{697}$. During the enumeration the first 80 million or more graphs the result would be identical, they give the same $\ell$ value, because we rewire only the last edge and it does not cause any change in $\ell$. 

We enumerated all the 10-node graphs with density 0.2 and it took a week CPU time to perform this task. This is "only" 886 163 135 graphs. Let us imagine how much time it takes to enumerate all the 100-nodes graphs with density 0.1.

\end{document}